\documentstyle[aps,twocolumn,epsf]{revtex}

\begin{document}

\draft

\title{Classical wave-optics analogy of quantum information processing}

\author{Robert J.C. Spreeuw}
\address{Van der Waals-Zeeman Instituut, Universiteit van Amsterdam,\\
Valckenierstraat 65, 1018 XE Amsterdam, the Netherlands\\
http://www.science.uva.nl/research/aplp}

\date{\today}

\maketitle

\begin{abstract}

An analogous model system for quantum information processing is
discussed, based on classical wave optics. The model system is applied
to three examples that involve three qubits: ({\em i}) three-particle
Greenberger-Horne-Zeilinger entanglement, ({\em ii}) quantum
teleportation, and ({\em iii}) a simple quantum error correction
network. It is found that the model system can successfully simulate
most features of entanglement, but fails to simulate quantum
nonlocality. Investigations of how far the classical simulation can be
pushed show that {\em quantum nonlocality} is the essential ingredient
of a quantum computer, even more so than entanglement. The well known
problem of exponential resources required for a classical simulation of
a quantum computer, is also linked to the nonlocal nature of
entanglement, rather than to the nonfactorizability of the state vector.

\end{abstract}


\narrowtext 

\section{Introduction}

The processing of quantum information is under intense investigation
because it enables applications which are either impossible or much less
efficient without the use of quantum mechanics
\cite{QFey82,QDeu85,QBenDiV00,BBouEkeZei00}. For example, secret keys
for encrypted communication can be distributed in ways that do not allow
an eavesdropper to gain information about the key
\cite{QBenBra84,QEke91,QTitBreZbi00,QNaiPetWhi00,QJenSimWei00}. A
formidable challenge still remains in the development of a universal
quantum computer, which should be able to solve certain problems in a
polynomial time where a classical computer requires exponential time. 

In this paper we explore an analogy of quantum information processing,
based on classical (optical) waves
\cite{QSpr98,QCerAdaKwi98,QKwiMitSch00}. This approach is inspired by
the observation that some of the essential properties of quantum
information are in fact wave properties, where the wave need not be a
quantum wave. By exploring the limits to where the classical analogy can
be pushed, we aim to obtain information about the subtle but profound
differences between the quantum system and the classical wave system. In
addition, the classical systems may serve as model systems for the
corresponding quantum systems, e.g. elucidating the mathematical
structure of a problem. 

Quantum bits, or ``qubits'' \cite{QSch95} are different from classical
bits in several important ways. A first important difference is that
qubits can exist in a superposition of the two binary states
$\{|0\rangle,|1\rangle\}$. A second crucial difference is that qubits
can be entangled. Superpositions are of course well known for classical
waves, so they are not exclusively quantum mechanical. Entanglement, on
the other hand, is commonly regarded as a quintessential quantum
phenomenon. It typically plays a role whenever quantum physics defies
``common sense'' and produces counterintuitive effects. Famous examples
are Schr\"odinger's cat \cite{QSch35,MonMeeKin96,BruHagDre96}, the
Einstein-Podolsky-Rosen paradox \cite{QEinPodRos35}, Bell's
inequality \cite{QBel65,AspGraRog82}, and, more recently, in quantum
cryptography
\cite{QBenBra84,QEke91,QTitBreZbi00,QNaiPetWhi00,QJenSimWei00},
teleportation \cite{QBenBraCre93,QBouPanMat97,QBosBraDeM98,QFurSorBra98}
and quantum computation \cite{QDeu85,QBenDiV00}. 

A classical analogy of entanglement has previously been constructed on
the basis of classical (light) waves \cite{QSpr98,QCerAdaKwi98}. The
analogy captures most features typically associated with entanglement,
such as a nonfactorizable state vector. However, the analogy fails to
produce effects of quantum nonlocality, thus signaling a profound
difference between two types of entanglement: ({\em i}) ``true,''
multiparticle entanglement and ({\em ii}) a weaker form of entanglement
between different degrees of freedom of a single particle. Although
these two types look deceptively similar in many respects, only type
({\em i}) can yield nonlocal correlations. Only the type ({\em ii})
entanglement has a classical analogy. 

In this paper the analogy is applied to concepts from quantum
information processing, such as elementary quantum gates and simple
quantum networks. We extend previous work on optical analogies,
concentrating here in particular on three-bit examples. We explore how
far we can push the classical analogy and to what extent the classical
system may be useful. 

We start in Sec.~\ref{sec:cebits} by briefly reviewing the concepts
outlined in Ref.~\cite{QSpr98}, introducing the ``cebit'' as the
classical counterpart of the qubit. The next three sections,
\ref{sec:ghz}--\ref{sec:errorcorrection}, are each devoted to one
specific application of the classical analogy. In Sec.~\ref{sec:ghz}
three-cebit entanglement, or so-called Greenberger-Horne-Zeilinger
states, is discussed. In Sec.~\ref{sec:teleportation} the teleportation
of a cebit is discussed. In Sec.~\ref{sec:errorcorrection} a simple
error correction network is described, which can correct either bit
flips or phase errors. Finally, in Sec.~\ref{sec:scaling} we discuss how
the resources required by the analogy scale with the number of cebits.
Conclusions are given in Sec.~\ref{sec:conclusion}.

\section{A classical Hilbert space of cebits}
\label{sec:cebits}

We briefly review the classical analogy as described by Spreeuw
\cite{QSpr98} and which is closely similar to that described by Cerf
{\em et al.} \cite{QCerAdaKwi98}. The state vector of a qubit,
$|\psi\rangle=q_0 |0\rangle+q_1 |1\rangle$, is specified by two complex
probability amplitudes, $(q_0, q_1)$. We replace these by two complex
classical wave amplitudes, the argument of the complex amplitudes
representing the phase. The resulting two-component complex vector
$(c_0, c_1)$ is the classical counterpart of a qubit and will be called
a {\em cebit}. The amplitudes $(c_0, c_1)$ may be macroscopic
and can in principle be measured directly. It should be noted that the
choice of optical waves is not essential for the analogy. Other
classical waves such as sound, water waves, or even coupled pendula
could be used in principle. 

We write the cebits in a notation which is a slightly modified version
of the familiar bra-ket notation of quantum mechanics. We use
parentheses for the cebits, instead of brackets, so that it is always evident 
whether we are
dealing with cebits or qubits. Similar to the bra and ket pair,
we use a {\em parent} $(\theta|$ and {\em thesis} $|\theta)$, which are
Hermitian conjugate to each other,
\begin{eqnarray}
  |\theta) & = & c_0 |0)+c_1 |1) \\
  (\theta| & = & c^\ast_0 (0|+c^\ast_1 (1|
\end{eqnarray}
The cebits form a Hilbert space where the Hermite product is given by
``parentheses,'' e.g. $(\theta|\theta)=|c_0|^2+|c_1|^2$.

As an example of a cebit, we can take the pair of complex amplitudes
describing the horizontal and vertical polarization components of a
laser beam. This two-component complex vector is known as a Jones vector
\cite{Jon41}. Here we will call it the {\em polarization cebit}.
Alternatively we can take the complex amplitudes of two spatially
separate laser beams (spatial modes). The two amplitudes now being
associated with position rather than polarization, we will call this
pair a {\em position cebit}.

\subsection{Measurements and unitary operations on a cebit}

\begin{figure}
\noindent
\centerline{\epsfxsize=80mm\epsffile{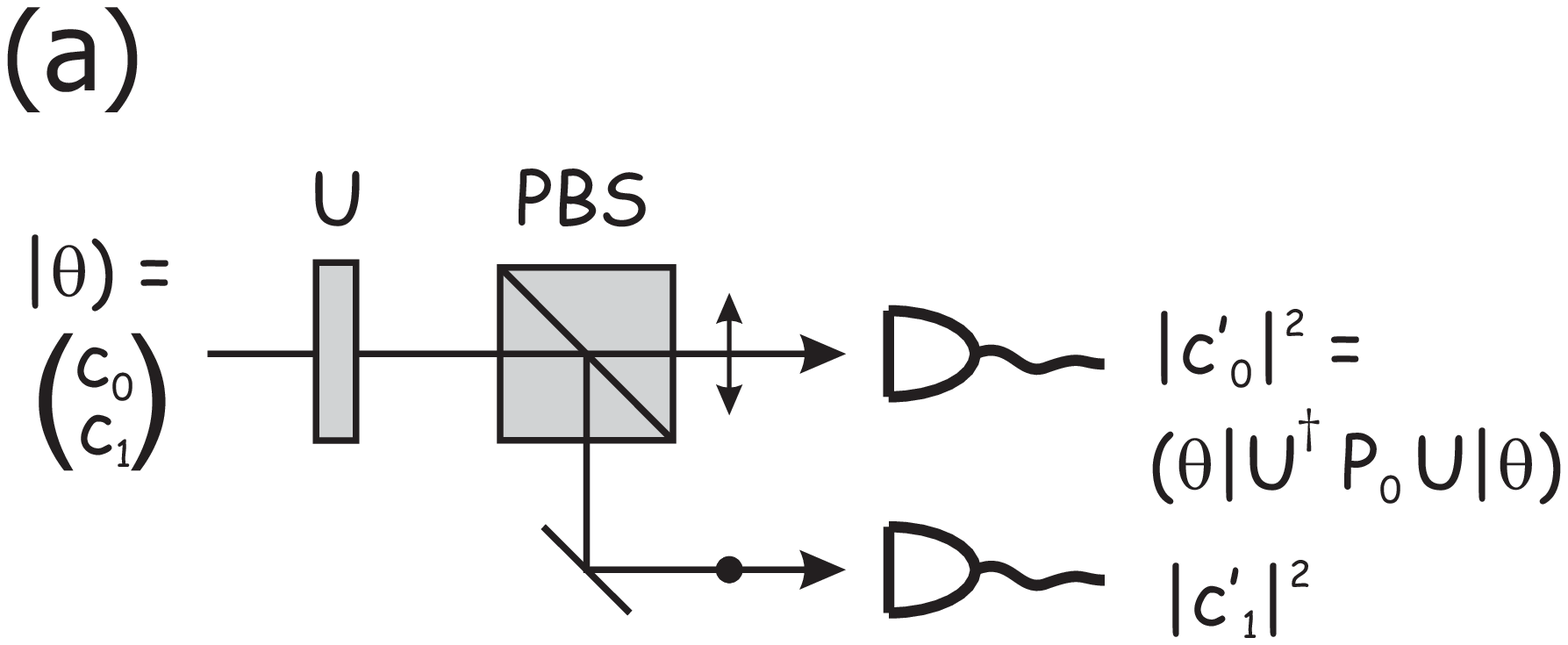}}
\vspace*{0.2cm}
\centerline{\epsfxsize=80mm\epsffile{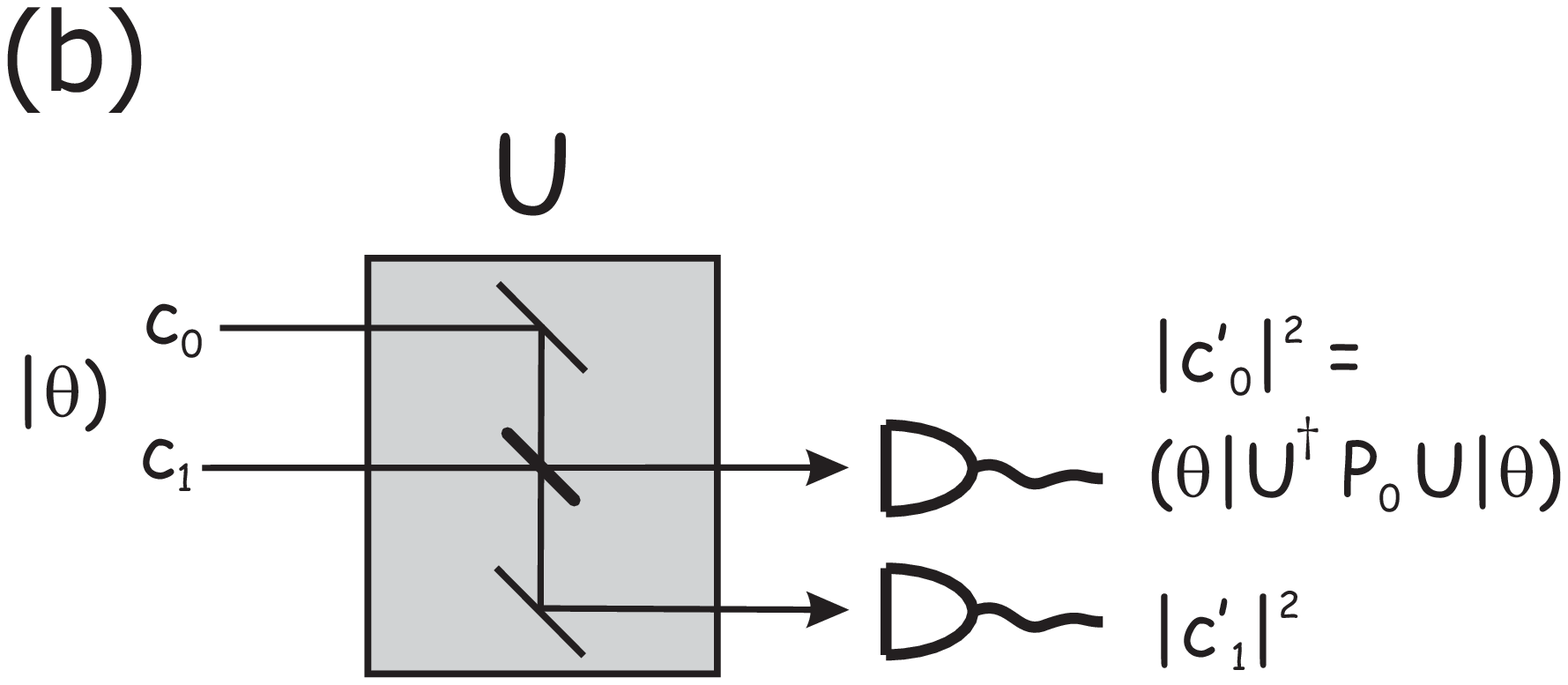}}
\vspace*{0.2cm}
\caption{(a) Measurement of a polarization cebit $|\theta)$ using a
polarizing beamsplitter PBS and photodetectors. The signals are
``expectation values'' of projection operators, $P_i=|i)(i|$. Measurements
in a different basis can be performed by first performing a unitary
operation $U$ on the polarization. (b) Measurement of a position cebit.
In this case the unitary operation consists of a beamsplitter with
variable splitting ratio and phase. 
\label{fig:cebitmeas}}
\end{figure}

Measurements on a cebit can be performed using photodetectors. For the
measurement of a polarization cebit we place photodetectors at the
outputs of a polarizing beam splitter, which transmits the horizontal
and reflects the vertical component, see Fig.~\ref{fig:cebitmeas}(a).
Identifying the transmitted and reflected components with the amplitudes
$c_0$ and $c_1$, respectively, the photodetector signals are
proportional to $|c_0|^2$ and $|c_1|^2$.

Instead of the intensities, one could in principle also measure the
amplitude and phase of the classical wave. Depending on the frequency,
the complex amplitudes could be measured either directly, or by mixing
them with a strong local oscillator in a homo- or heterodyne
measurement. Here we restrict ourselves to intensity measurements, which
also give access to complete information about the complex amplitudes.
The reason is that one can first produce two identical copies of a
classical light beam using a beam splitter, and perform measurements on
the copies in different bases of measurement. With qubits this is of
course impossible, since they cannot be copied. 

A measurement can be performed in a different basis by first
performing a unitary operation on the cebit, $|\theta)\rightarrow
|\theta')=U|\theta)$. Writing $(c'_0, c'_1)$ for the amplitudes of 
$|\theta')$, the measured signals are
\begin{equation}
  |c'_i|^2=(\theta'| P_i |\theta' )=(\theta| U^{\dag} P_i U |\theta ),
\end{equation}
where $P_i=|i)(i|$ are projection operators. For the polarization cebit
the transformation $U$ is performed using polarization optics, such as
quarter-wave plates (QWP), polarization rotators, etc. For example, a
Hadamard gate can be realized by a half-wave plate (HWP), its fast axis
oriented at $22.5^\circ$ with respect to the vertical direction, see also 
Fig.~\ref{fig:gates}(a). More
general, a universal SU(2) operator can be realized by a sequence of
three rotatable retarders, QWP-HWP-QWP \cite{Bha89}. 

For a measurement on a position cebit we simply place photodetectors in
each beam of the beam pair to measure two intensities proportional to
$|c_0|^2$ and $|c_1|^2$. A different measurement basis can again be
obtained by first performing a unitary operation, see
Fig.~\ref{fig:cebitmeas}(b). This operation should now mix the two
spatially separated amplitudes, using beam splitters. The Hadamard gate
can be realized by a 50/50 beam splitter, with properly adjusted phases
of the inputs and outputs, see Fig.~\ref{fig:gates}(b). A universal
SU(2) operator can be realized by a Mach-Zehnder interferometer,
consisting of two 50/50 beam splitters and three adjustable phase
delays: in one of the inputs, in one of the interferometer arms, and in
one of the outputs \cite{YurMcCKla86}. 
In fact it has been shown that any unitary $N\times N$ matrix can be 
realized as an optical multiport \cite{QRecZeiBer94}.

\subsection{Multiple cebits}

We can generalize this procedure and construct the Hilbert space of
multiple cebits, which should be spanned by the tensor products of basis
states of the Hilbert spaces of the individual cebits. We concentrate
here on a system of three cebits. Since the quantum state of three qubits is
described by eight probability amplitudes, 
\begin{equation}
  |\Psi\rangle=q_{000}|000\rangle+q_{001}|001\rangle+\ldots+q_{111}|111\rangle,
\end{equation}
the corresponding cebit version should also contain
eight amplitudes. This can be accomplished using four laser beams 
with polarization. 

\begin{figure}
\noindent
\centerline{\epsfxsize=80mm\epsffile{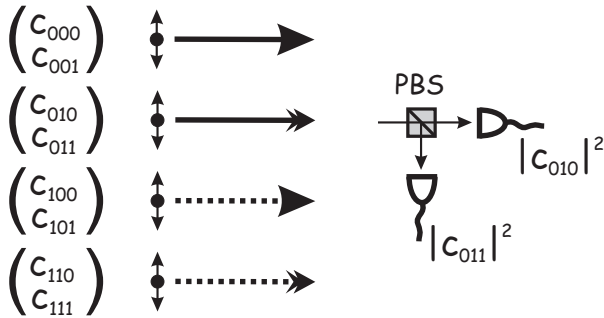}}
\vspace*{0.2cm}
\caption{Eight classical wave amplitudes of four polarized beams,
together encoding three cebits. The first cebit is here depicted by line
style (solid for 0, dotted for 1), the second by arrow style (single for
0, double for 1), the third by polarization (vertical for 0, horizontal
for 1).
\label{fig:threecebits}}
\end{figure}

We label the amplitudes $c_{000}\ldots c_{111}$ as
shown in Fig.~\ref{fig:threecebits}. Each pair of amplitudes
$(c_{ij0},c_{ij1})$ constitutes the Jones
polarization vector of one of the four beams. One might thus easily,
though wrongly, identify each of the four Jones vectors with one cebit.
In fact the four Jones vectors constitute only three cebits. Four
amplitudes are associated with any bit value of a given cebit. For
example, the subspace where the polarization cebit has value 0 is
specified by four amplitudes that vanish, $c_{ij1}=0$.

The remaining two cebits are position cebits. The ``most significant
cebit'' (MSC), associated with the first index $i$ of $c_{ijk}$,
describes coarse position, bit value 0 meaning that the lower beam pair
is dark. The middle bit $j$ describes the fine position, within a pair
of beams, bit value 0 now meaning that the lower beam within each beam
pair is dark. In Fig.~\ref{fig:threecebits} the bit values for the MSC
are indicated by line style (solid vs. dashed), for the middle cebit by
arrow head style (single vs. double) and for the least significant cebit
by polarization direction (arrow vs. dot.)

\section{Greenberger-Horne-Zeilinger states}
\label{sec:ghz}

As a first example of three-bit states we construct the optical
analog of the three-particle entangled state introduced by Greenberger,
Horne, and Zeilinger (GHZ) \cite{QGreHorZei89,QGreHorShi90},
\begin{equation}
  |\Psi_{\rm GHZ}\rangle = 
  \frac{1}{\sqrt{2}}\left( |000\rangle + |111\rangle \right) .
\end{equation}

Let us first briefly recall the remarkable  properties of this quantum
state. These 
become apparent by measuring the four different observables $\sigma_x^2
\sigma_y^1 \sigma_y^0$, $\sigma_y^2 \sigma_x^1 \sigma_y^0$, $\sigma_y^2
\sigma_y^1 \sigma_x^0$, and $\sigma_x^2 \sigma_x^1 \sigma_x^0$, (in
short: ``$xyy$'', ``$yxy$'', ``$yyx$'', and ``$xxx$'', respectively).
Here, $\sigma_j^i\,(j=x,y,z)$ are the Pauli matrices for qubit $i$, 
\begin{equation}
  \sigma_x=
   \left(\begin{array}{cc}
    0 & 1 \\
    1 & 0
   \end{array}\right),\,
  \sigma_y=
   \left(\begin{array}{cc}
    0 & -i \\
    i & 0
   \end{array}\right),\,
  \sigma_z=
   \left(\begin{array}{cc}
    1 & 0 \\
    0 & -1
   \end{array}\right)
\end{equation}

The first three measurements, on $xyy$, $yxy$, and $yyx$, all yield the
value $-1$, since $|\Psi_{\rm GHZ}\rangle$ is an eigenstate with
eigenvalue $-1$. We could now attempt to predict the outcome of the
fourth measurement, on $xxx$, by assigning values $+1$ or $-1$ to the
individual spin components. For example, we could assign
$\sigma_x^2=\sigma_y^1=\sigma_x^0=+1$ and
$\sigma_y^2=\sigma_x^1=\sigma_y^0=-1$. This would predict a value $-1$
for the $xxx$ measurement. There are eight possible combinations of such
assignments, predicting unanimously the value $-1$ for $xxx$. On the
other hand, the quantum prediction is $+1$, since $|\Psi_{\rm
GHZ}\rangle$ is an eigenstate of $\sigma_x^2 \sigma_x^1 \sigma_x^0$ with
eigenvalue $+1$. The quantum prediction has recently been confirmed
experimentally, providing a dramatic demonstration of quantum
nonlocality \cite{QBouPanDan99,QPanBouDan00}. 

The optical version, $|\Theta_{\rm GHZ})$, consists of two laser beams
with orthogonal polarization (plus two beams which are dark), see
Fig.~\ref{fig:ghz}. We now pose the obvious question of what will be the
outcome of the equivalent cebit measurements for this ``classical GHZ
state''? We thus have to perform joint measurements on three cebits.

\subsection{Joint cebit measurements on the GHZ state}

For any three-cebit state $|\theta)$ a measurement on the cebits can be
performed by placing polarizing beam splitters (PBS) into each beam plus
a photodetector at each output, as shown in Fig.~\ref{fig:threecebits}.
This makes eight photodetectors in total, corresponding to the eight
basis states of the three-cebit Hilbert space. The signal of a
particular photodetector is proportional to $|c_{ijk}|^2$. Note that any
given detector is associated with three bit values at once, one for each
cebit. The choice of the measurement basis is again done by a unitary
operation on the three-cebit state. For example this may consist of a
sequence of three unitary operations for each cebit separately, as in
Fig.~\ref{fig:cebitmeas}.

\begin{figure}
\noindent
\centerline{\epsfxsize=80mm\epsffile{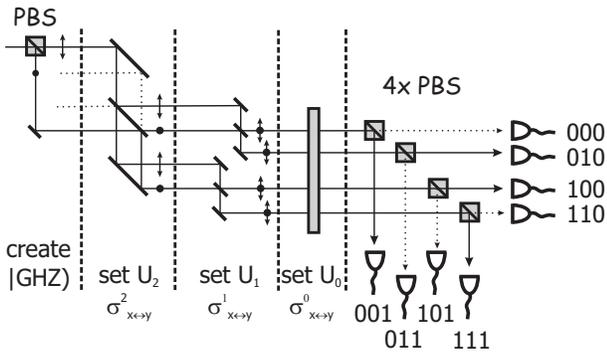}}
\vspace*{0.2cm}
\caption{Interferometer representing measurements on a three-cebit
entangled ``GHZ'' state. For each cebit $i$, the unitary operation $U_i$
sets the measurement basis to $x$ or $y$ (in spin-1/2 space). For $U_{1,2}$ this is done by
adjusting the phases of the mirrors and beam splitters. For $U_0$ the
polarization element is switched between a $45^\circ$ rotator ($x$
basis) and a quarter-wave plate oriented at $45^\circ$ ($y$ basis). The
solid and dashed lines identify the bright and dark outputs for either of
the settings $xyy$, $yxy$, or $yyx$. For the setting $xxx$ the bright ports
become dark and vice versa. Note that the $U_i$ commute, so their order
is arbitrary. 
\label{fig:ghz}} 
\end{figure}

The GHZ measurement thus translates into a classical interferometer,
shown in Fig.~\ref{fig:ghz}. It is straightforward to calculate the
signals that will be measured at the eight output ports of this
interferometer. If we set the measurement basis to $xyy$, we find that
four out of eight outputs are dark and that the remaining four have
equal intensity. The dark outputs are those labeled by 000, 011, 101,
and 110. The remaining, bright, outputs are just the ones corresponding
to a measurement result $-1$, if we identify bit values 0 and 1 by spin
values $+1$ and $-1$, respectively. 

The fact that four outputs carry equal intensity shows that the GHZ
state is not an eigenstate of any individual cebit operator. Instead, it
is an eigenstate of the three-cebit observable $xyy$. In other words, we
have not obtained a definite value for any cebit, but instead found that
the three cebits are correlated. If we change the measurement basis to
$yxy$ or $yyx$, we find that the same four outputs are dark and the same
four are bright. So again the measurement result is $-1$, as formulated
in GHZ language. The fact that each of the three measurements yields
four dark and four bright outputs is a striking demonstration of the
failure of any attempts to assign definite values $+1$ or $-1$ to the
individual cebit components.

Finally, we change the measurement basis to $xxx$ and find that in this
case the four dark outputs have now become the bright ones and vice
versa. Apparently the measurement result is now $+1$, in complete
correspondence with the quantum result for the real GHZ experiment.

\subsection{Discussion}

The cebit version of the GHZ experiment thus shows that in general it is
impossible to assign definite values to different degrees of freedom of
a single particle. The cebit analogy is described by the same
mathematics and may thus be used to elucidate the mathematical structure
of the problem. On the other hand, the underlying physics is different
in a very subtle way. 

The measurements on different degrees of freedom (cebits) of a single
particle are correlated, but the correlations will never be of a
nonlocal nature. The three cebits are by construction always chained
together. It is impossible to spatially separate them and perform
separate measurements on the three cebits in different locations.
Therefore no statements about local realism can be made on the basis of
the cebit experiment. 

Since the three-cebit GHZ state is a classical state of light, it would
of course be possible to make three identical copies and send these to
different locations to be analyzed. For example, one could measure
$\sigma_x^2$ in location 2, $\sigma_y^1$ in location 1, and $\sigma_y^0$
in location 0. However there will be no correlations between the
measurements in different locations. The correlations exist only between
the degrees of freedom of each local copy, not between different copies.

Nevertheless it is remarkable that the classical analogy reproduces
exactly the quantum correlations, since in the quantum case the nonlocal
correlations are commonly associated with the collapse of the
wavefunction due to a measurement. In the classical case,
the correlations are due to classical wave interference. A choice of
measurement basis leads to destructive interference in some output
channels, which then appears as correlations of cebit values. 

Finally, it should be noted that this GHZ experiment with cebits has not
actually been performed experimentally. However, we have described
essentially an interferometer for classical light, so there is no reason
to doubt these predictions.

\section{Teleportation of a cebit}
\label{sec:teleportation}

\subsection{Teleportation of a qubit}

Quantum teleportation allows the transmission of the quantum state of a
qubit between a sender and a receiver, usually called Alice and Bob.
Alice destroys the qubit by making a measurement, gaining however no
information on the qubit. The result of the measurement enables Bob to
recreate the qubit. We follow here the concept as introduced by Bennett
{\em et al.} \cite{QBenBraCre93}, which has recently been realized
experimentally by Bouwmeester {\em et al.} \cite{QBouPanMat97}.
Alternative teleportation schemes have also been demonstrated
\cite{QBosBraDeM98,QFurSorBra98}. The Rome experiment
\cite{QBosBraDeM98} may in fact be considered as a hybrid version, using
in addition to the entanglement between two separate photons, also the
different degrees of freedom of one photon. For a detailed comparison
between the different teleportation schemes, see Ref.
\cite{BBouEkeZei00}. Briefly, the Bennett procedure works as follows. 

First Alice and Bob share an EPR pair of qubits, in the entangled state
$|\Psi_{\rm EPR}\rangle=(|01\rangle-|10\rangle)/\sqrt{2}$. Alice now
performs a Bell state measurement on her part of the EPR pair plus the
qubit whose state $|\psi\rangle$ she wants to teleport. This means
she performs a measurement on the two-qubit state in the basis of the 
four Bell states: 
\begin{eqnarray}
  |\Psi^\pm\rangle & = & \frac{1}{\sqrt{2}} 
    \left( |01\rangle\pm |10\rangle \right) \\ 
  |\Phi^\pm\rangle & = & \frac{1}{\sqrt{2}} 
    \left( |00\rangle\pm |11\rangle \right) 
\end{eqnarray} 
Alice then informs Bob which of the four Bell states she found, by
sending him two bits of classical information. Finally, Bob performs one
out of four unitary operations on his half of the EPR pair, following
the two-bit instruction received from Alice. This leaves Bob's qubit
in the state $|\psi\rangle$.

\subsection{Optical implementation}

\begin{figure}
\noindent
\centerline{\epsfxsize=80mm\epsffile{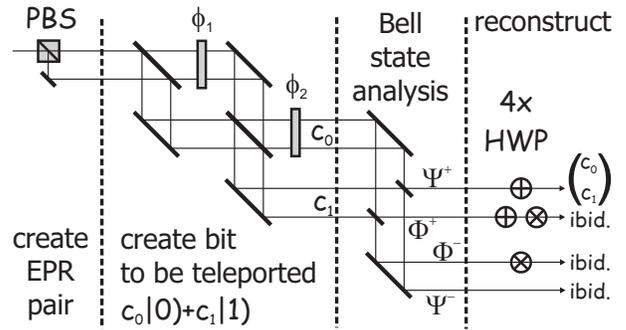}}
\vspace*{0.2cm}
\caption{Optical network to ``teleport'' a position cebit to the
polarization cebit. In the final state all four beams have the same
polarization, with coefficients $(c_0,c_1)$ equal to that of the
position cebit created earlier. 
\label{fig:teleport}} 
\end{figure}

The optical analogy of this protocol is sketched in
Fig.~\ref{fig:teleport}. A somewhat different implementation has been
described in Ref.~\cite{QCerAdaKwi98}. The implementation discussed here
starts by generating the ``EPR pair,'' a two-cebit state $|\Theta_{\rm
EPR})=[|01)-|10)]/\sqrt{2}$ of entangled position and polarization. This
is done using a polarizing beam splitter. In the next step an additional
position cebit $|\theta_2)$ is created. This is the cebit that is to be
``teleported''. This additional cebit is created by splitting the EPR
pair into two copies with relative amplitudes $c_0$ and $c_1$. This is
done using a Mach-Zehnder interferometer consisting of two 50/50 beam
splitters and two adjustable phase delays, $\phi_1$ in one of the arms
and $\phi_2$ in one of the output ports. The to-be-teleported cebit,
$|\theta_2)=c_0|0)+c_1|1)$, apart from an overall phase, is selected by
setting the two phase delays $\phi_{1,2}$. 

If we want to teleport position cebit $|\theta_2)$ into the
polarization cebit, we must now perform a Bell state analysis on the two
position cebits. In fact it is not necessary to do any photodetection.
It is sufficient to perform a basis transformation to the Bell basis,
using a set of beam splitters. These mix the $00$ with the $11$
amplitudes and the $01$ with the $10$ amplitudes. After this unitary
transformation, the four beams correspond to the four Bell states, i.e.
the four possible outcomes of Alice's measurements.

Finally, depending on the Bell state, one out of four operations is
performed on the polarization cebit. This means that a different
polarization operation is performed on each of the four beams. The
operations are the identity and three different spin rotations by
$\pi/2$ about the axes $x$, $y$, and $z$. The optical implementations
for the three rotations are respectively: a QWP with its
axes rotated by $\pi/4$, a $\pi/2$ optical rotator, and a QWP
with its fast axis in the vertical direction. Note that in
Fig.~\ref{fig:teleport} the rotator has been replaced by two
HWP's with a relative orientation of $45^\circ$.

This optical system teleports the position cebit $|\theta_2)$ into the
polarization cebit. This means that now the four beams carry equal
polarization, for every possible setting of the phase delays
$\phi_{1,2}$ of the Mach-Zehnder interferometer. Furthermore the
polarization is given by the Jones vector $(c_0,c_1)$, as determined by
the phase delays.

\subsection{Discussion}

By transforming to the Bell basis and post-processing the polarization,
the three-cebit state $[c_0|0)+c_1|1)]\otimes |\Theta_{\rm EPR})$ has
been transformed into the three-cebit state $|\theta_{21})\otimes
[c_0|0)+c_1|1)]$. Here we denote by $|\theta_{21})$ the final state of
the two position cebits, 2 and 1. If desired, the two position cebits
contained in $|\theta_{21})$ can now be discarded. The four beams have
well defined phase relationships, which do not depend on $(c_0,c_1)$.
Therefore the four beams can be combined into a single beam using beam
splitters. 

In the end, what we have accomplished is to combine the amplitudes of
two different beams into the polarization components of a single beam.
Of course, this is a rather trivial optical task that could have been
accomplished in a much simpler way. However, by following the
teleportation protocol, several subtle differences become clear between
qubits and cebits. 

In the quantum version, the Bell state analysis seems impossible without
a nonlinear process, i.e. interaction between the qubits
\cite{QKwiWei98}. In contrast, in the cebit version, the Bell state
analysis is easily performed using passive linear components. Again,
like in the GHZ example, the inseparability of the cebits in space poses
a limitation. Rather than teleporting a qubit state over some distance
in space, the state of one cebit is transferred to another cebit. In
this case a position cebit was ``teleported'' to the polarization cebit.
However, the source and target cebits are necessarily always part of the
same composite multiple-beam system. Bob's half of the EPR pair and
Alice's classical information cannot be transmitted separately.

\section{Error correction networks}
\label{sec:errorcorrection}

As a third application of the optical analogy of qubits, we consider a
simple error correction network \cite{QSho95,QSte96}, correcting either
for bit flip errors or phase errors. In principle, a network that
corrects both bit flips and phase flips may also be constructed
optically. However, this would require at least five cebits
\cite{QLafMiqPaz96,QBenDiVSmo96}, i.e. 16 beams of polarized light. A
quantum logic network that corrects bit flips is shown in
Fig.~\ref{fig:errcorr}(a). The correction network for phase errors is
obtained by applying a Hadamard gate to each qubit, both before and
after the bit flip region. The network makes use of controlled NOT
(c-NOT) gates and a Toffoli gate. These are described first.

\subsection{Controlled NOT and Toffoli gates}

The c-NOT gate for cebits takes several shapes, depending on which cebit
is the control bit and which the target. For simplicity, we describe
here the c-NOT gates for the situation of two cebits, one position and
one polarization cebit. 

\begin{figure}
\noindent
\centerline{\epsfxsize=70mm\epsffile{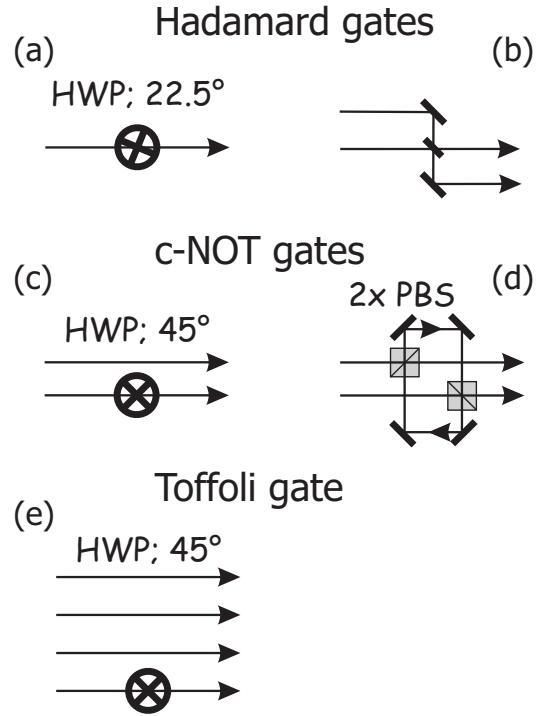}}
\vspace*{0.2cm}
\caption{Examples of elementary gates for cebits. (a) Hadamard gate for
a polarization cebit, consisting of a HWP oriented at $22.5^\circ$. (b)
Hadamard gate for a position cebit, consisting of a 50/50 beamsplitter.
(c) c-NOT gate where the position cebit is the control, the polarization
the target. The polarization bit is flipped by means of a half-wave
plate (HWP). (d) c-NOT gate, polarization controls position.
Polarizing beam splitters (PBS) reroute the beams in such a way that the
two beams exchange position, conditioned on the polarization. (e)
Toffoli gate, flipping the polarization cebit, conditioned on two
position cebits. 
\label{fig:gates}}
\end{figure}

The simplest situation occurs when the position is the control cebit and
the polarization the target. The gate should thus flip the polarization
if the position cebit has the value 1, i.e. if the light is in the lower
beam. The polarization flip between horizontal and vertical is obtained
using a HWP oriented at $45^\circ$ with respect to
vertical. We simply place the HWP in the lower beam to
obtain a c-NOT gate, see Fig.~\ref{fig:gates}(c). 

By a simple extension a Toffoli gate is obtained, where both control
cebits are position cebits, the target polarization. In this case we
have four beams and the subspace where both position cebits have value 1
is formed by the lowermost beam of the four. Therefore the Toffoli gate
is implemented by placing the half-wave plate in the lower beam, see
Fig.~\ref{fig:gates}(e). 

If the roles of target and control bits in the c-NOT gate are reversed,
polarization becoming the control and position the target bit, the c-NOT
gate becomes somewhat more complicated. Conditioned on the polarization,
the gate should reverse the upper and lower beams. This can be
accomplished using the optical arrangement shown in
Fig.~\ref{fig:gates}(d). 

\subsection{Cebit-flip error correction network}

\begin{figure}
\noindent
\centerline{\epsfxsize=80mm\epsffile{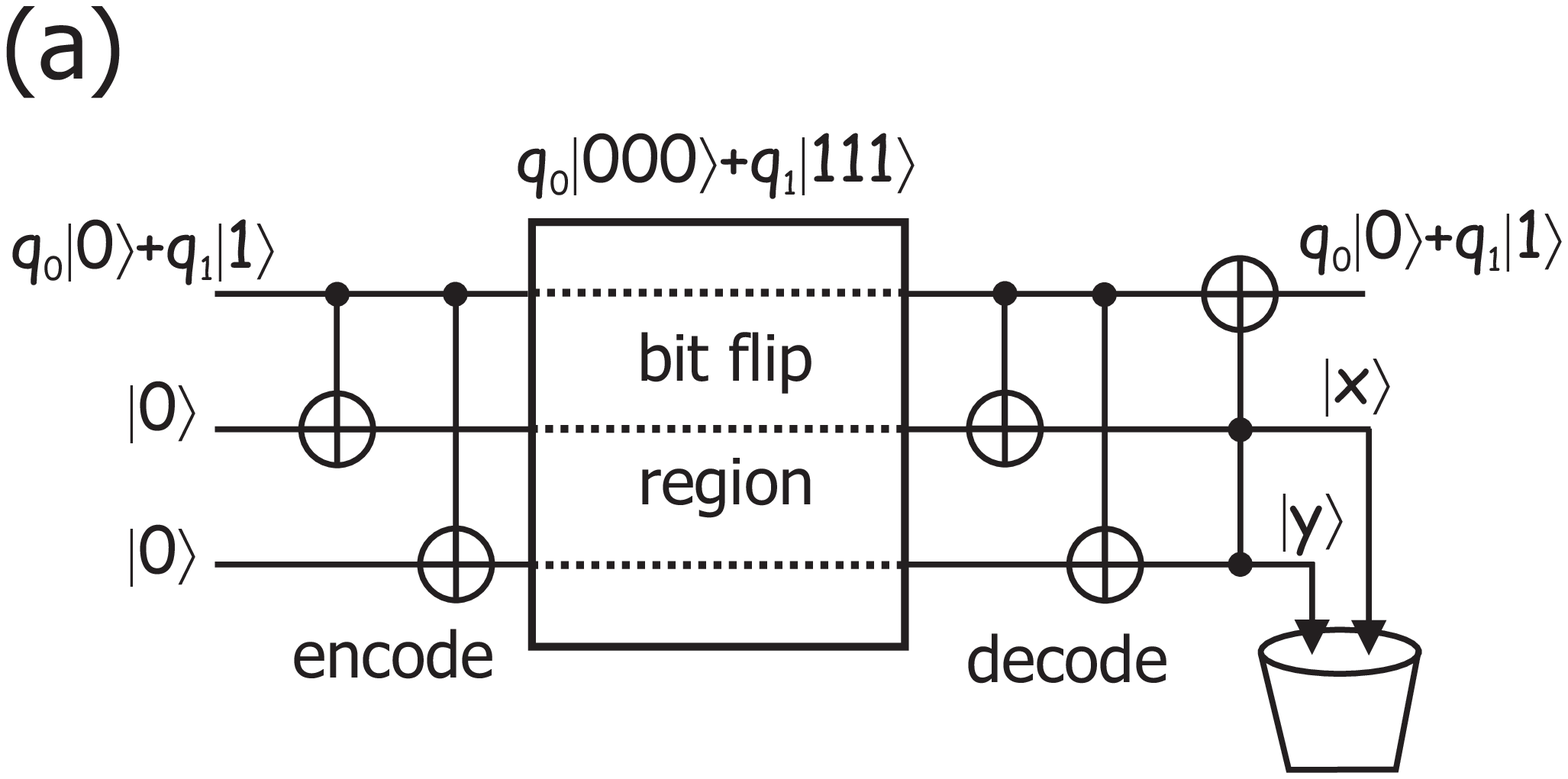}}
\vspace*{0.2cm}
\centerline{\epsfxsize=80mm\epsffile{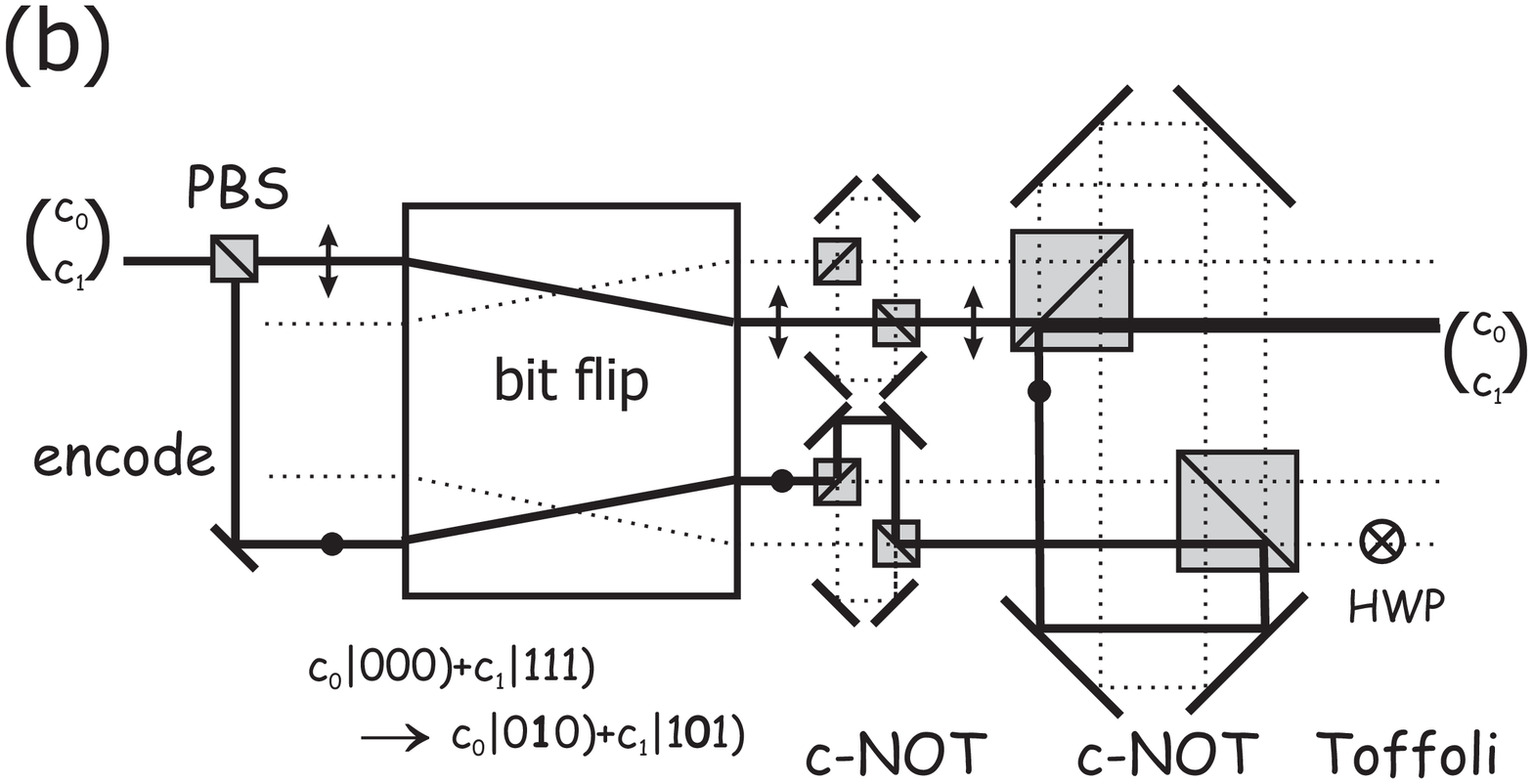}}
\vspace*{0.2cm}
\caption{(a) Quantum logic network correcting for bit flip errors. (b)
The cebit version. The incident polarization is restored and comes out
in one of the four output beams, depending on which cebit was flipped.
The solid line traces the optical beam path in the case that the middle
cebit is flipped. Note that networks correcting for phase errors may be 
obtained by applying a Hadamard gate on each qubit/cebit, both before 
and after the bit flip region.
\label{fig:errcorr}}
\end{figure}

The optical cebit version of the bit-flip correction network is shown in
Fig.~\ref{fig:errcorr}(b). In the qubit version,
Fig.~\ref{fig:errcorr}(a), the first two c-NOT gates serve to store a
single qubit into a three-cebit entangled state:
$q_0|0\rangle+q_1|1\rangle\rightarrow q_0|000\rangle+q_1|111\rangle$.

In the optical cebit version this can be accomplished using a single
polarizing beam splitter (PBS), acting on the incident polarization
cebit $c_0|0)+c_1|1)$. The component $c_1|1)$ is split off and moved
into the fourth beam, whereas the component $c_0|0)$ forms the first
beam. Thus we obtain the three-cebit state $c_0|000)+c_1|111)$. The next
section in the optical diagram is the region where possible bit flips
can occur. The bit flip is corrected using two c-NOT gates and a Toffoli
gate. The c-NOT gates are a generalized version of those shown in
Fig.~\ref{fig:gates}(d). Both have the polarization as the control cebit
and a position cebit as the target. Finally, the Toffoli gate is formed
by placing a half-wave plate in the fourth beam, just as in
Fig.~\ref{fig:gates}(e). 

Let us now suppose that the middle cebit flips,
\begin{equation}
  c_0|000)+c_1|111)\rightarrow c_0|010)+c_1|101).
\end{equation}
Optically, this bit flip entails swapping the beams within each pair,
i.e. swapping beams 1 and 2 and likewise beams 3 and 4, see
Fig.~\ref{fig:errcorr}(b). In the same Figure we also see how
the network corrects the error, by tracing the optical path through the
sequence of polarizing beam splitters. Eventually, the two components
are recombined so that the original input polarization $c_0|0)+c_1|1)$
is restored. The port where this restored polarization exits (i.e.
the values of the two position cebits) is determined by which cebit was
erroneously flipped. 

Just like in the qubit version, the entire three-bit state at the output
of the error-correction network is uncertain. However, unlike in the
qubit version, the two cebits whose state is uncertain cannot easily be
discarded. Whereas in the qubit version the two unused bits are separate
particles and can be removed, this is no longer possible in the cebit
version. Removing the two uncertain cebits would mean recombining the
four beams into one. This cannot be done without loss of intensity,
because the relative phases and amplitudes are uncertain. The fact that
we cannot discard the two unknown cebits can be seen as another failure
to simulate the nonlocality in the quantum system. Again this feature is
lost in the translation to the cebit version.

\section{Scaling and resources}
\label{sec:scaling}

The optical analogy discussed in this paper is based on replacing $n$
two-state particles by a single particle (photon) with $n$ two-state
degrees of freedom. Since in both cases we deal with a $2^n$-dimensional
Hilbert space, the two situations may seem rather similar. However the
difference is profound. It is well known that for this simulation
procedure one pays the price of exponential resources
\cite{QFey82,QEkeJoz98}. Here, each extra cebit doubles the required number of
light beams. The amount of optical hardware required also grows
exponentially with the number of cebits. 

This difference in scaling behavior poses more than merely a technical
problem. One can easily see that this is in fact a dramatic limitation
by examining how far the classical system could be scaled up. We can
attempt to stack light beams as densely as possible. The cross section
of a single beam is limited by optical diffraction to the order of the
square of the wavelength, or $\lambda^2\sim 10^{-12}$~m$^2$ in the
optical region. In a stretch of the imagination, let us assume that the
stack of beams can be made to fill the entire universe, with a diameter
$D\sim 10^{26}$~m. The number of beams can then be increased to
approximately $(D/\lambda)^2\sim 10^{64}$. Although this is a huge
number, the number of cebits (using two polarizations) is still only
$n=1+\log_2 10^{64}\approx 214$. Even if we use a wavelength
$\lambda\sim 10^{-10}$~m, the number of cebits is no larger than 240.

Another resource that may pose a limitation is the total optical power.
Since the light is distributed over an exponentially growing number of
light beams, the intensity per beam decreases exponentially. This
becomes important if one is interested to eavesdrop on the information
processing, i.e. if one wishes to measure the intermediate state. This
possibility should in principle exist in a classical system. However, as
a result of the exponential scaling behavior the possibility to
eavesdrop on the classical system is lost. 

Thus the most important difference between qubits and cebits becomes
clear by looking at the scaling behavior while increasing the number of
bits. Obviously, the amount of required resources like beam splitters
and such will also increase exponentially and quickly deplete the
available matter in the known universe.

\section{Conclusion}
\label{sec:conclusion}

In summary, the optical analogy of quantum information, as introduced in
Ref.~\cite{QSpr98} has been applied to three examples involving three
qubits/cebits. 

The three-cebit ``GHZ'' entangled state shows that it is impossible to
assign values $\pm 1$ to the $x$ and $y$ component of all three
cebits, in such a way that all joint measurements are correctly
predicted. The predicted results are formally identical to the quantum
predictions for the qubit version of the experiment. Thus the optical
analogy is a useful tool to visualize and elucidate the mathematical
structure of the problem. The crucial difference is that the three
cebits cannot be spatially separated, so that the optical system cannot
be used for testing local realism. 

Like in the GHZ example, in the example of teleportation the
unseparability of the cebits in space again poses a limitation. Rather
than teleporting a qubit state over a distance in space, the state of
one cebit is transferred to another cebit. In the example, a position
cebit was ``teleported'' to the polarization cebit. However, the source
and target cebits are always part of the same composite multiple-beam
system. 

In the third example we considered a simple error correction network
that corrects for either bit flips or phase errors. The limitation
encountered here was that the extra two cebits required for the
correction network cannot be easily disposed of after finishing the
protocol. This limitation can again be seen as a consequence of the lack
of nonlocality. 

The analogy is based on replacing the $2^n$-dimensional Hilbert space of
$n$ two-state particles (qubits) by the $2^n$-dimensional Hilbert space
of a single particle (photon) with $n$ two-state degrees of freedom. It
is well known that by doing so, one pays the price of exponential
resources. In the present examples the number of light beams and optical
components grows exponentially with the number of cebits. This scaling
behavior has been previously been attributed to a consequence of
entanglement \cite{QEkeJoz98}. 

Nevertheless, the classical analogy is able to simulate most features of
entanglement, in particular the nonfactorizability of the state vector.
Nonlocality, on the other hand, has defied all attempts at classical
simulation so far. The limitations of the simulations encountered in the
examples above, could be traced back to a lack of (quantum) nonlocality.
It seems therefore that it is the {\em nonlocal nature} of the
entanglement which is intimately linked to the exponential scaling
behavior. 

Apart from the scaling issue, it would superficially seem that one could 
build a quantum computer without the need for nonlocality. After all,
if one is not interested in sending qubits/cebits to spatially separate
locations, one could hope to perform the computation locally. 
However, even in the simplest error correction network as discussed above, 
the classical simulation encountered the limitation that the unused bits
could not easily be disposed of. This can again be seen as a consequence
of a lack of nonlocality in the classical case. The debate about whether
entanglement is the essential ingredient of a quantum computer is still
going on. The classical simulations as discussed here strongly suggest
that, {\em quantum nonlocality} is the essential ingredient of a
quantum computer, even more so than entanglement.

\section*{Acknowledgments}

This research has been made possible by a fellowship of the Royal
Netherlands Academy of Arts and Sciences.


\begin{thebibliography}{10}

\bibitem{QFey82}
R.~P. Feynman, Int. J. Theor. Phys. {\bf 21},  467  (1982).

\bibitem{QDeu85}
D. Deutsch, Proc. R. Soc. Lond. A {\bf 400},  97  (1985).

\bibitem{QBenDiV00}
C.~H. Bennett and D.~P. {DiVincenzo}, Nature {\bf 404},  247  (2000).

\bibitem{BBouEkeZei00}
D. Bouwmeester, A.~K. Ekert, and A. Zeilinger (Eds.), 
{\em The Physics of Quantum Information} (Springer, Berlin, 2000).

\bibitem{QBenBra84}
C.~H. Bennett and G. Brassard,  in {\em Proc. {IEEE} Int. Conference on
  Computers, Systems and Signal Processing, Bangalore, India} (IEEE, New York,
  1984), p.\ 175;
  C.~H. Bennett, G. Brassard, and A. Ekert, Sci. Am. {\bf 267},  26  (1992).

\bibitem{QEke91}
A.~K. Ekert, Phys. Rev. Lett. {\bf 67},  661  (1991).

\bibitem{QTitBreZbi00}
W. Tittel, J. Brendel, H. Zbinden, and N. Gisin, Phys. Rev. Lett. {\bf 84},
  4737  (2000).

\bibitem{QNaiPetWhi00}
D.~S. Naik {\it et~al.}, Phys. Rev. Lett. {\bf 84},  4733  (2000).

\bibitem{QJenSimWei00}
T. Jennewein {\it et~al.}, Phys. Rev. Lett. {\bf 84},  4729  (2000).

\bibitem{QSpr98}
R.~J.~C. Spreeuw, Found. of Phys. {\bf 28},  361  (1998).

\bibitem{QCerAdaKwi98}
N.~J. Cerf, C. Adami, and P.~G. Kwiat, Phys. Rev. A {\bf 57},  R1477  (1998).

\bibitem{QKwiMitSch00}
P. Kwiat, J. Mitchell, P. Schwindt, and A. White, J. Mod. Opt. {\bf 47},  257
  (2000).

\bibitem{QSch95}
B. Schumacher, Phys. Rev. A {\bf 51},  2738  (1995).

\bibitem{QSch35}
E. Schr{\"o}dinger, Naturwiss. {\bf 23},  807,823,844  (1935).

\bibitem{MonMeeKin96}
C. Monroe, D. Meekhof, B. King, and D. Wineland, Science {\bf 272},  1131
  (1996).

\bibitem{BruHagDre96}
M. Brune {\it et~al.}, Phys. Rev. Lett. {\bf 77},  4887  (1996).

\bibitem{QEinPodRos35}
A. Einstein, B. Podolsky, and N. Rosen, Phys. Rev. {\bf 47},  777  (1935).

\bibitem{QBel65}
J. Bell, Physics (N.Y.) {\bf 1},  195  (1965).

\bibitem{AspGraRog82}
A. Aspect, P. Grangier, and G. Roger, Phys. Rev. Lett. {\bf 49},  91  (1982).

\bibitem{QBenBraCre93}
C.~H. Bennett {\it et~al.}, Phys. Rev. Lett. {\bf 70},  1895  (1993).

\bibitem{QBouPanMat97}
D. Bouwmeester {\it et~al.}, Nature {\bf 390},  575  (1997).

\bibitem{QBosBraDeM98}
D. Boschi {\it et~al.}, Phys. Rev. Lett. {\bf 80},  1121  (1998).

\bibitem{QFurSorBra98}
A. Furusawa {\it et~al.}, Science {\bf 282},  706  (1998).

\bibitem{Jon41}
R.~C. Jones, J. Opt. Soc. Am. {\bf 31},  488  (1941).

\bibitem{Bha89}
R. Bhandari, Phys. Lett.~A {\bf 138},  469  (1989).

\bibitem{YurMcCKla86}
B. Yurke, S.~L. {McCall}, and J.~R. Klauder, Phys. Rev.~A {\bf 33}, 4033
(1986).

\bibitem{QRecZeiBer94}
M. Reck, A. Zeilinger, H.~J. Bernstein, and P. Bertani, Phys. Rev. Lett. {\bf
  73},  58  (1994).

\bibitem{QGreHorZei89}
D. Greenberger, M. Horne, and A. Zeilinger,  in {\em {Bell}'s Theorem, Quantum
  Theory, and Conceptions of the Universe}, edited by M. Kafatos (Kluwer
  Academic, Dordrecht, 1989), p.\ 73.

\bibitem{QGreHorShi90}
D. Greenberger, M. Horne, A. Shimony, and A. Zeilinger, Am. J. Phys. {\bf 58},
  1131  (1990).

\bibitem{QBouPanDan99}
D. Bouwmeester {\it et~al.}, Phys. Rev. Lett. {\bf 82},  1345  (1999).

\bibitem{QPanBouDan00}
J.-W. Pan {\it et~al.}, Nature {\bf 403},  515  (2000).

\bibitem{QKwiWei98}
P.~G. Kwiat and H. Weinfurter, Phys. Rev. A {\bf 58},  R2623  (1998).

\bibitem{QSho95}
P.~W. Shor, Phys. Rev. A {\bf 52},  R2493  (1995).

\bibitem{QSte96}
A.~M. Steane, Proc. R. Soc. London A {\bf 452},  2551  (1996).

\bibitem{QLafMiqPaz96}
R. Laflamme, C. Miquel, J.~P. Paz, and W.~H. Zurek, Phys. Rev. Lett. {\bf 77},
  198  (1996).

\bibitem{QBenDiVSmo96}
C.~H. Bennett, D.~P. {DiVincenzo}, J.~A. Smolin, and W.~K. Wootters, Phys. Rev.
  A {\bf 54},  3824  (1996).

\bibitem{QEkeJoz98}
A. Ekert and R. Jozsa, Phil. Trans. R. Soc. Lond. A {\bf 356},  1769  (1998).

\end{thebibliography}

\end{document}